\PassOptionsToPackage{table}{xcolor}
\documentclass[sigconf,review]{acmart}
\usepackage{tikz}
\usetikzlibrary{shapes.geometric, arrows.meta, positioning, calc, shadows}
\usepackage{enumitem}
\setlist{topsep=0pt,leftmargin=*}

\copyrightyear{2026}
\acmYear{2026}
\setcopyright{cc}
\setcctype{by}
\acmConference[ICSE-FoSE '26]{2026 IEEE/ACM 48th International Conference on Software Engineering: Future of Software Engineering }{April 12--18, 2026}{Rio de Janeiro, Brazil}
\acmBooktitle{2026 IEEE/ACM 48th International Conference on Software Engineering: Future of Software Engineering (ICSE-FoSE '26), April 12--18, 2026, Rio de Janeiro, Brazil}
\acmPrice{}
\acmDOI{10.1145/3793657.3793874}
\acmISBN{979-8-4007-2479-4/2026/04}

\makeatletter
\patchcmd{\maketitle}{\@copyrightspace}{}{}{}
\begin{document}

\title{SE Journals in 2036: Looking Back at the Future We Need  to Have}
\author{Tim Menzies \textsuperscript{1},
Paris Avgeriou \textsuperscript{2}, 
Robert Feldt \textsuperscript{3}, 
Mauro Pezzè \textsuperscript{4}, \\
Abhik Roychoudhury \textsuperscript{5},
Miroslaw Staron \textsuperscript{6}, 
Sebastian Uchitel \textsuperscript{7},
Thomas Zimmermann \textsuperscript{8}}

\affiliation{%
  \institution{\small
    \textsuperscript{1}NC State University, USA \quad
  \textsuperscript{2}University of Groningen, the Netherlands \quad
    \textsuperscript{3}Chalmers University, Sweden \quad
    \textsuperscript{4}USI, Switzerland
  }
   \city{}
  \country{}
}

\affiliation{%
  \institution{\small
      \textsuperscript{5}National University of Singapore \quad  
    \textsuperscript{6}Univesity of Gothenburg, Sweden \quad
    \textsuperscript{7}Imperial College, UK \quad
    \textsuperscript{8}UC Irvine, USA
  }
   \city{}
  \country{}
}
\email{timm@ieee.org ,  p.avgeriou@rug.nl , robert.feldt@chalmers.se , mauro.pezze@usi.ch}
\email{abhik@nus.edu.sg , miroslaw.staron@chalmers.se ,  sebastian.uchitel@gmail.com , tzimmer@uci.edu}

\renewcommand{\shortauthors}{A redline of editors.}


\begin{abstract}
In 2025, SE publishing faces an existential crisis 
of scalability. As our communities swell globally and integrate 
fast-moving methodologies like LLMs, traditional peer-review 
practices are   collapsing under the strain. The "bureaucratic 
anomaly" of monolithic review has become mathematically unsustainable, 
creating a stochastic "lottery" that punishes novelty and exhausts 
researchers.

This paper, written from the perspective of 2036, documents potential 
solutions. Here,   the editors of ASE, EMSE, IST, 
JSS, TOSEM and TSE   dream a collective 
dream of a brighter future. In summary
first we stopped fighting (The Journal  Alliance). Then we fixed the process (The Lottery / Unbundling / Fixing the Benchmark Graveyard). And then we fixed the culture (Cathedrals/Bazaars).

\end{abstract}

\maketitle

\pagestyle{plain}
\section{Houston, We have a Problem}
\begin{figure*}
\includegraphics[width=.7\linewidth]{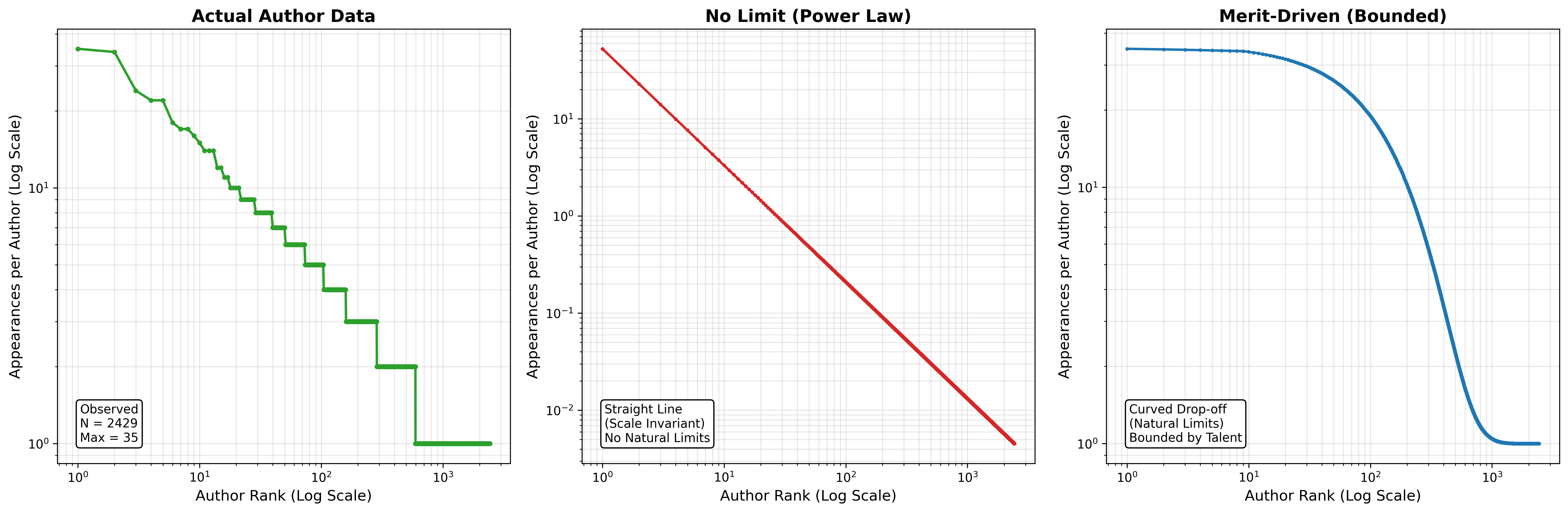}
\caption{ 
Consider two citation models.
(a) \textbf{Stochastic:} Feedback loops cause unbounded citations; 
i.e. authors with many citations always earn yet more citations due 
to their viability. This would appear as a straight line on log-log 
graph of ``paper rank'' versus ``citations standing''.
(b) \textbf{Merit-based:} Quality has some natural limits where most 
author quality clusters around a mean, with a few large outliers. 
Citations from such a population should appear like a bell curve, 
which on a log-log graph would be a non-linearity.
The Green plot (left) shows data on authors whose papers scored ten 
most citations in leading SE forums 2013 to 
2023~\cite{monperrus2025most}. Note that the SE data matches a 
power law distribution (center) better than a merit-driven curve 
(right).}\label{curves}
\end{figure*}

 Researchers
can be surprised by the differences between
the craft of research and the mechanics of publication.
\begin{itemize}
\item
To the individual, research is often solitary (or small team) task,
that is a quality-driven pursuit—an intense scrutiny of specific 
ideas where success is measured by depth of insight and/or breadth 
or impact.
\item
Publication, on the other hand, is a complex process defined, at least to some extent, by 
scale. Experiments at AI conferences~\cite{
cortes2021inconsistencyconferencepeerreview} reveal that
independent review teams disagree on 50\% of accepted papers. This
suggests that peer review and ``getting published'' are stochastic rather 
than a precise measure. Our own analysis (shown in 
Figure~\ref{curves}) showed that just like AI, SE reviewing is more 
stochastic-based than merit-based. 
\end{itemize}

As the editors of some high volume SE journals (ASE, EMSE, 
IST, JSS, TOSEM, TSE) we view this landscape through the lens of Price’s 
Law~\cite{Price1963}; i.e. 50\% of scientific 
contributions are generated by the square root of the research 
population ($\sqrt{N}$). This leads to  a  logistical 
imperative: we need methodologies to rapidly sample the total 
population ($N$) to identify those core contributors. 

But our duty extends beyond simply harvesting the "breakthroughs." 
A healthy journal ecosystem must also support the foundational 
trial-and-error—the "negative" results and iterative attempts—that 
allow the high-impact work to emerge. We must efficiently filter 
for the signal, without discarding the necessary noise of the 
scientific process.

A 2025 editorial in TOSEM~\cite{roychoudhury2025year} underscored 
the large volume of submissions. In terms of concrete numbers, there 
were around 2000 submissions to TOSEM in 2025 including both first 
submissions and revisions. These submissions were handled within 67 
days on average, which does put a high load on the entire reviewer 
community. Associate editors can send up to 20 invitations to
secure reviewers. These bottlenecks can cause delays for certain 
papers. JSS decided in 2025 to establish a cap of 20 review 
invitations per submission. If this limit was reached, the 
submission was assumed to not fit the scope of the journal, and was 
desk-rejected. While this draconian measure may seem unfair, it is 
one way to address wasting too much time for the authors and editors.

When researchers and practitioners turn to arXiv for real-time 
advances in LLMs, traditional forums are being left behind, 
rendered irrelevant. The community attempted to patch these leaks:
\begin{itemize}
\item
To ensure that good papers at journals were not ignored, the Journal First Initiative\footnote{Initially 
organized in 2016 by Matt Dwyer.} allowed authors of published 
journal papers to present those at major conferences. 
\item To tap the wealth of research that may be languishing in 
arxiv, the Journal Ahead Workshop (JAWs\footnote{Suggested by Mauro Pezze in 2025 and then conducted in 2026 jointly with TOSEM} was organized initially by TOSEM 
and TSE for the International Conference on Software Engineering 
(ICSE) 2026 (and then later, by other journals at other venues). The aim of the JAWs workshop was to invite early ideas  to be presented in a workshop of a major conference, subsequent to 
which selected papers may be submitted to a journal.
\end{itemize}

While timely, these initiatives were not enough. They did not 
address the core scalability problem, nor the friction caused by 
forcing fast-moving "Bazaar" artifacts into slow "Cathedral" review 
cycles. To achieve true scalability, we needed to adopt an 
engineering perspective on the review process itself. What was its 
true function? Where, precisely, were we wasting resources?

Taking that first step required us to confront the central myth of 
our discipline: that peer review is a reliable filter through 
which specialists objectively identify quality. For a brief period, the 
community operated under the somewhat misguided assumption that the 
only path to scientific truth lay through a Byzantine architecture 
of rebuttals, revisions, and gate-keeping. By 2026 it was very 
clear that that kind of peer review had become mathematically 
unsustainable and empirically suspect.

We say this because, in any active research area, it is 
mathematically impossible for reviews from $N$ authors to be 
conducted by the $\sqrt{N}$ most active researchers.
\begin{itemize}[noitemsep, topsep=0pt, leftmargin=*]
\item
Table~\ref{corepeople} shows the number of authors of papers with 
``top 10 citations'' at leading SE venues from 
2013 to 2023 (from~\cite{monperrus2025most}). 
\item The table compares top-10 authors to a 
``back-of-the-envelope'' calculation for all the other authors at 
those venues. In this sample, the SE field has $\approx 19,000$ 
researchers publishing at senior venues, of which 
$\frac{500}{19,000}\approx 2.6\%$ are the most active 
{\em core}\footnote{This number is larger than what might be 
predicted by Price's law~\cite{Price1963} since 
$\sqrt{18,500}=136 < 500$. Nevertheless, it is still consistent 
with ``most of the highly recognized work comes from a very small 
core''.}.
\end{itemize}
This skew creates a critical asymmetry: the core cannot possibly 
review the massive volume of the ``long tail,'' turning the 
reviewer shortage into one of scale rather than quality. 
When the core does not review the rest of the field, potentially 
useful work is overlooked, and reviewing becomes a random process 
rather than a method for 
promoting new groundbreaking ideas with a short review turnaround time. There have been concerns as well about allocating reviews based on submissions. Since certain authors are very prolific - allocating reviews based on submission may cause certain authors to have a huge say in the paper acceptance. This remains a consideration.

\begin{table}[!t]
\begin{center}
{\footnotesize
\begin{tabular}{lccl} 
\textbf{Group} & \textbf{$N$} & \textbf{Cites} & \textbf{Assumptions} \\\hline
\textbf{Top-10 Authors} & 450--550 & 150--1000+ & $\mu=6$--$10$ papers/author \\
\textbf{Other Authors} & $\sim$18,000 & 5--20 & $\mu=2.5$ papers/author  
\end{tabular}
}
\end{center}
\caption{SE has a small core of most cited researchers.}\label{corepeople}
\end{table}

This paper outlines how we escaped that randomness. Writing from the 
perspective of 2036, we document the "Great Restructuring" of the 
last decade. Faced with the collapse of the traditional model, the 
community stopped trying to patch the leak and instead redesigned 
the pipe. What follows are the six pivotal reforms—ranging from 
political alliances to automated infrastructure—that transitioned 
software engineering from a cottage industry into a scalable 
scientific discipline.

\section{Six Fixes for SE Journals}

 In summary
first we stopped fighting (The Journal Alliance). Then we fixed
the process (The Lottery / Unbundling / Fixing the Benchmark
Graveyard). And then we fixed the culture (Cathedrals/Bazaars).

\subsection{The SE Journal Alliance}
Before we could fix the mechanics of peer review, we had to fix the 
structure of our venues.
By 2025, the competition  between venues had  increased 
inefficiencies in an already stretched community. As noted by 
Uchitel, the overlaps between venues were significant enough that 
manuscripts could be sent almost anywhere.
This led to the wasteful practice of "forum shopping," where authors 
of a rejected paper would send their work to a different venue with 
minimal changes, hoping for a better stochastic outcome. The effort 
in reviewing the original manuscript was completely wasted:
\begin{itemize}
    \item There was no incentive for transparency on  manuscript's 
    history.
    \item Anonymous reviews were not shared between venues.
    \item Organizers had no access to reviewer names to request 
    re-reviews.
\end{itemize}
While manuscripts jumped from venue to venue, reviewers did not. 
Resources were tied to specific venues, treating the global reviewer 
pool as fragmented silos.

To solve this collective-action failure, the top SE journals formed 
the \textbf{SE Journal Alliance} in 2026.
To solve the collective-action failure of reviewer burnout, the top SE journals formed an informal alliance in 2026. This group established shared norms for reviewer load and, crucially, a portable credit layer. 

This alliance shifted us  from competition 
to collaboration, sharing procedures, infrastructure, 
reviewers, and historical data.

\begin{itemize}
    \item \textbf{Portable Reviews:} In an accepted paper today, one 
    can see the full lineage of its evolution across the Alliance. 
    Previous reviews accompany the paper, preventing the loss of 
    valid critiques and stopping the "reset" button of forum shopping.
    
    \item \textbf{Shared Credit \& Norms:} The Alliance established 
    shared norms for reviewer load and a "portable credit layer." 
  This system tracks editorial and review contributions across all member venues, converting "invisible labor" into durable academic capital. By making this data visible to hiring and tenure committees, the community transformed peer review from a volunteer burden into a recognized metric of professional standing. 
\end{itemize}

In a sense, venues today are less "paper-evaluation-machines" and 
more like digests with specific viewpoints, selecting from a shared, 
community-wide pool of evaluated research.

\subsection{The Lottery}
Given the randomness reported above, it was natural to 
take the advice of Roumbanis~\cite{roumbanis2019lottery} and 
restructure journal reviewing as a lottery process. 
 Papers were scored by very fast moving pre-review teams (previously
 known as  the ``desk reject team''). The higher the score,
 the more likely a paper was selected for a traditional  human review.
 All papers that scored over a certain threshold were always reviewed.
 Others took their chances with the lottery. By adjusting this threshold according to the number of available reviewers, journals could control how many reviews were needed. 

  Crucially, because the lottery reduced the total volume, we could 
afford to abandon the "one-shot" review model. It was replaced by a 
\textbf{continuous dialog process}.
The "one-shot" review model, which often amplified misunderstanding and noise, was replaced by a continuous dialogue process. Every paper passing the initial gates triggers a private discussion page where authors, reviewers, and editors engage in structured Q\&A. This interaction improves calibration and allows for the "meritorious acceleration" of high-potential work. Post-decision, these dialogues are published alongside the paper (with protected reviewer identities), serving as a vital educational resource and a record of the paper’s evolution.

Another aspect that helped  was a pre-review process - as done in journals like Nature - where only     promising papers are sent out for peer review. TOSEM journal already started this process in 2025, with three levels of screening before a paper is sent for peer-revuew, by Editor in Chief, Senior Associate Editor and Associate Editor.

\subsection{Review  = Automatic + Manual}

To manage this volume without surrendering quality, we adopted the 
"Human+AI" philosophy proposed by Manrai et 
al.~\cite{manrai2025accelerating}. They argued that AI should never be 
an arbiter of scientific merit, but serves as essential infrastructure.

We therefore moved away from "AI scoring" (which implies judgment) to 
"AI Triage" (which implies compliance). When reviewers look at a paper, 
they are checking for many items, some of which are purely structural 
and can be detected automatically.

Some automatic support proved useful for this triage process. 
The authors of this paper  do not endorse fully   automatic triage of academic papers. Nevertheless, when reviewers look at a paper, they are checking for many things, some of which can be detected automatically.
For example:
 \begin{itemize}[noitemsep, topsep=0pt, leftmargin=*]
 \item
In  the Ralph Empirical Standards initiative~\cite{ralph2021empiricalstandardssoftwareengineering}, SE researchers from around the world have commented on the features they expect in dozens of different  kinds of SE papers.
\item
These documents mention  around 135  review items. After asking Gemini to abstract those items, we found that there are variants of 12 core features.  Significantly for our discussion here,  some of these could be reviewed either by automatic methods
or humans glancing very quickly at a  paper. 
\item
We hence
 used a hierarchical triage system where h  human review
is only triggered if lower-level requirements are met.
\end{itemize}
By 2028, Open Science became a mandatory functional requirement to scale "Structural Gates," replacing static PDFs with a "Compilation Model." Submissions must now "compile"—executing analysis scripts in standardized containers—before human review, with exceptions only for justified sensitive industrial data. This shift enabled the AI revolution by allowing agents to automatically verify reproducibility and statistical consistency, freeing experts to focus entirely on interpretation, novelty, and high-level logic rather than broken technical details
(i.e. where the underlying mechanics—data, code, or statistics—functionally fail or contain errors).

After that shift, here is our review plan.

{\bf Phase 1: Structural Gates (Automated)}
The entry level consists of binary checks. These are low-cost, structural validations. Failure here indicates an immature draft or policy violation.
\begin{itemize}
    \item \textbf{Hard Exclusions:} Fail if clearly unrelated domain, not scientific manuscript, or violates alliance policy.
    \item \textbf{Structural Checks:} Presence of Related Work, Threats/Limitations, Results vs. Discussion distinction, and Replication Artifacts (code/data links).
\end{itemize}

{\bf Phase 2: Descriptive Gates (Extraction)}
This tier scans for definitions and risks. It creates a ``Scope Profile'' to simplify human review (Fast/Medium speed).
\begin{itemize}
    \item \textbf{Scope Profile:} Extract claimed contribution type, objects of study, evaluation mode, and keywords/taxonomy mapping.
    \item \textbf{Risk Flags:} Flag mismatches to venue expectations or misalignment between claimed and AI-generated descriptions (non-failing).
    \item \textbf{Essential Definitions:} Verification of stated research goals, context, and variables.
\end{itemize}

{\bf Phase 3: Cognitive Review (Human Expert)}
Only papers passing prior gates reach experts. This phase focuses on logic, validity, and flag resolution (Slow/Deep speed).
\begin{itemize}
    \item \textbf{Scope Adjudication:} Decide in/out of scope with rationale based on Phase 2 flags. May judge borderline/novel work requiring reframing. 
    \item \textbf{Deep Review:} Assess motivation, sampling logic, methodology, and analysis. Double check  the  Related Work for discussion of relevant connections
    \item \textbf{Evaluation:} Rigor of validation. For experiments, standard metrics; for vision papers:   review of counter-arguments.
\end{itemize}
While controversial,   triage   eliminated the incredibly time-consuming ``hair-splitting'' required to rank top-tier papers. It reduced the adversarial nature of reviews. However, it was only partially adopted. By 2030, many conferences used a lottery for the ``middle tier'' of papers, drastically reducing the debate time during PC discussions.

One unexpected benefit of the lottery was that it enabled a wider range of artifacts to be published. For example, GitHub repositories with clear README files could be parsed and scored automatically, allowing review teams to evaluate well-motivated code with illustrative examples like any other research output.

\subsection{ Cathedrals and Bazaars}
The structural changes described above would have failed without a parallel shift in the culture of incentives. By 2026, it was clear that the "one size fits all" publication model was forcing two distinct scientific species into the same cage.

To resolve the crisis, we formalized a "Two-Speed" culture, effectively recognizing that a healthy ecosystem needs both \emph{Cathedrals} and \emph{Bazaars} (otherwise known as   ``deep science'' and "agile science'').
 For the foundational work, in
   deep science "Cathedral" papers\footnote{
   Adapted from \url{https://compsci.science/slowscience}.}
   authors explored topics in extensive depth:
   \begin{itemize}
    \item \textbf{Deep PhD:} In cathedral science, the standard PhD thesis in Software Engineering shifted to include only one scientific publication. However, this single contribution was required to be of very high depth and rigor.
    \item \textbf{Hiring Metrics:} Hiring committees and tenure boards ceased "paper counting." Instead, they adopted the "Rule of Three": candidates were evaluated solely on the qualitative assessment of their three most significant publications. We note that certain countries already follow this practice. 
\end{itemize}
This restored the "velocity of depth," allowing reviewers to scrutinize fewer submissions with significantly greater diligence.

\begin{figure}[!t]
\includegraphics[width=3.5in]{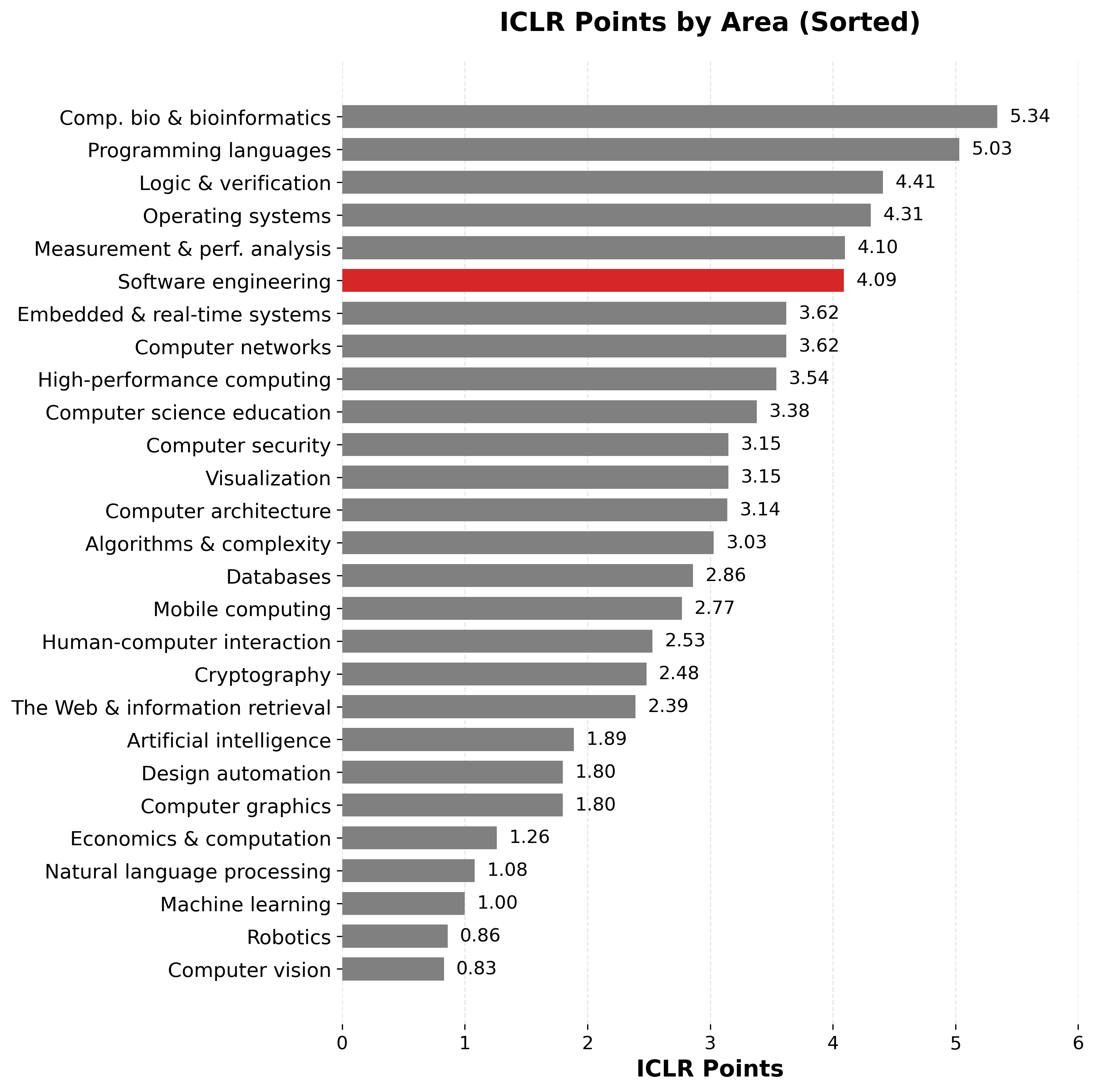}
\caption{
SE author  density is four times that of machine learning
(from~\cite{luo2025iclr}) reflecting either higher effort is required or vastly different authorship cultures.}\label{iclr}
\end{figure}

Simultaneously, we stopped treating velocity as a vice. We acknowledged that   "Bazaar" agile scientists
(often highly practitioner-focused, exploring  rapid tool developments and  empirical snapshots) were being suffocated by the multi-year  paper cycles. Conversely, when these rapid inputs were forced into the slow lane, they clogged the review machinery for everyone else.
In SE,  clogging is painfully apparent.
In a recent post, Luo~\cite{luo2025iclr} offers a heuristic ranking for how many
people have to work together to write a paper in different fields.
After normalizing the effort for machine learning papers to one, Luo
ranks the effort associated with other fields (see
Figure~\ref{iclr}). Note that the author density required
for SE papers was much higher than most other areas;
i.e.,
there was ample space to redesign what we mean as a ``paper''.

It was  realized that "Bazaar" research needed a different validity construct entirely—one based on utility and immediacy rather than archival permanence. The specific mechanisms we engineered to support this fast stream are detailed in the next section.

\begin{figure}[!t]
\begin{center}
\includegraphics[width=3.5in]{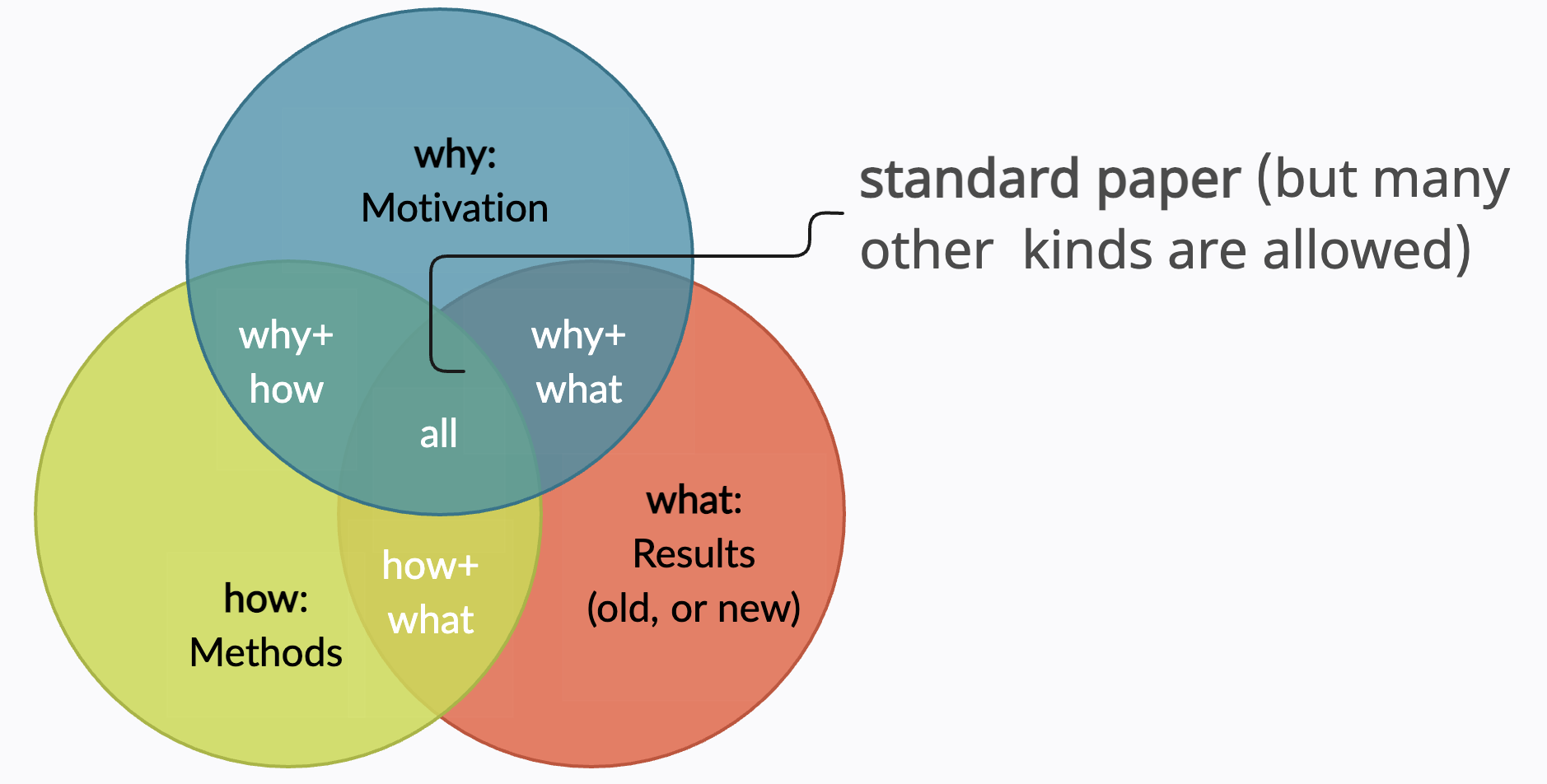}
\end{center}
\caption{Different kinds of papers.}\label{types}
\end{figure}

\subsection{Unbundling the Monolith}
As shown in Figure~\ref{types},  standard technical papers are a  container 
of \emph{motivation}, \emph{method}, and \emph{results}. But by 
2030, we realized that forcing every contribution to carry all three 
containers was inefficient.

 For the deep science track, we allowed researchers to publish modular 
units (Micro-publications):
\begin{itemize}
    \item A \textbf{Vision Statement} only needs motivation.
    \item A \textbf{Registered Report} contains just motivation and 
    method.
    \item A \textbf{Tools Paper} or \textbf{Literature Review} may be 
    just method.
    \item A \textbf{Replication Paper} might be just motivation and 
    results.
\end{itemize}

Prior to 2025, the SE community was already experimenting with 
dividing up papers in this way. For example, Ernst and 
Baldassarre~\cite{ernst2023registered} described \emph{registered 
reports} as scientific publications that begin with the peer review 
of a detailed research protocol before the study is conducted. In 
terms of Figure~\ref{types}, the protocol would be a publication of 
motivation and methods, and the subsequent results report would be 
mostly results.

Unbundling the "monolithic paper" into  smaller units allowed for 
 specialize reviewing. For example, a "Methodology" paper could be 
reviewed quickly by a statistics expert, without   evaluating
a year's worth of results simultaneously. By 2030, this reduced the 
average reading time per reviewer by 40\%, as they only had to 
critique one aspect of the research life cycle at a time.

For the faster agile track, we solved the "ivory tower" problem by detaching 
the practitioner payload from the scientific manuscript.
Journal papers are no longer considered by software practitioners as 
"write-only," inaccessible works. Instead, every accepted submission 
is now accompanied by {AI-generated Summaries and Videos}.
\begin{itemize}
    \item \textbf{Access:} This content is placed outside the 
    "paywalls" of traditional publishers. It requires no subscription, 
    allowing it to reach the global engineering workforce instantly.
    \item \textbf{Format:} These artifacts explain the context, 
    problem, and key findings in a way that is accessible and 
    time-efficient. Practitioners can digest the core contribution 
    in minutes.
    \item \textbf{Integration:} Artifacts like tools or datasets are 
    embedded directly into practitioner workflows via AI tooling 
    supports.
\end{itemize}
Crucially, these summaries are \emph{not} subject to pre-publication 
peer review. They are published independently. The quality control 
here is not "guarding," but usage. Practitioner feedback on these 
artifacts is tracked and—vital to the culture shift mentioned in 
Section 4—this feedback now counts towards meeting impact criteria 
in academic career paths.
 
\subsection{Escaping the Benchmark Graveyard}
By 2025, empirical software engineering faced a paradox. We had more 
data than ever, but less insight. The root cause was our relationship 
with benchmarks.

The 2026 "Moot" paper~\cite{moot2026benchmarks}
warned that
benchmarks can be  a \emph{catalyst} or a \emph{cage}. Used 
thoughtfully, they broaden the questions we ask. But used 
carelessly, they suffer a tragic life cycle:
\begin{enumerate}
    \item \textbf{Rejected:} "Find Data? That will never happen."
    \item \textbf{Respected:} "Fine... it helps sometimes."
    \item \textbf{Expected:} "You must compare to this baseline."
    \item \textbf{Exhausted:} The research graveyard where innovation 
    stalls; results become repetitive and derivative.
\end{enumerate}

 By the mid-20s, several SE sub-fields had   entered the "Exhausted" phase.
 The widespread reuse of open datasets had reduced 
validation to a narrow form of testing. Researchers were merely 
demonstrating that "the method works on the existing dataset" or "the 
tool is better by $x$\% than another tool."

This practice occurred without sufficient critical examination of the 
validity of the chosen dataset itself. We were optimizing for the 
cage, ignoring whether the method was effective in real-world 
settings or industrial environments. Consequently, impact 
stagnated—industry looked to non-peer-reviewed blogs because academic 
papers were winning games on stale scoreboards.

 To escape the graveyard, the Journal Alliance introduced the 
"Catalyst Criteria" in 2027. We  stopped rewarding papers 
that only offered minor percentage gains over existing baselines.
Instead, acceptance now requires work that actively expands the 
field; e.g.:
\begin{itemize}
    \item \textbf{New Tasks or Domains:} Moving beyond the "standard" 
    datasets to prove utility in messy, real-world contexts.
    \item \textbf{Challenging Assumptions:} Using benchmarks to 
    disprove prevailing wisdom (not just climbing a leaderboard).
    \item \textbf{Extending the Benchmark:} Contributing back to the 
    community by improving the data, tooling, or organization of the 
    benchmark itself.
\end{itemize}
This shift transformed benchmarks from a ceiling that limited us into 
a floor upon which we stood to reach higher.

\section{Conclusion}
Looking back from 2036, the crisis of the mid-20s appears not as a 
disaster, but as a necessary correction. The "bureaucratic anomaly" 
of 2025—where we tried to apply 19th-century reviewing methods to 
21st-century AI velocity—forced us to innovate.

The solution lay in humility. By admitting that peer review is 
stochastic, we adopted the Lottery. By admitting that PDFs are 
opaque, we demanded Executable Papers. By admitting that 
practitioners ignore paywalls, we built the Bazaar.

Ultimately, the Journal Alliance didn't just save our venues; it 
saved our time. We moved from a system of "Quality Control by 
Guarding" (where we spent 90\% of our energy filtering noise) to 
"Quality Control by Dialog" (where we spend 90\% of our energy 
improving signal). Software Engineering is no longer just a 
publishing machine; it is, once again, a scientific community.

\section*{Disclaimer}

This paper is co-authored by several editor-in-chiefs. The views expressed capture their individual experiences in the role, and do not necessarily capture the views of the professional societies for their corresponding journals.

\bibliographystyle{ACM}
\bibliography{refs}

\end{document}